# Oxygen vacancy formation energies in Sr-doped complex perovskites: ab initio thermodynamic study

Denis Gryaznov,*[1,4] Mike W. Finnis,[2] Robert Evarestov,[3] and Joachim Maier[1]

[1] Max Planck Institute for Solid State Research, Heisenbergstr. 1, D-70569 Stuttgart, Germany
[2] Department of Materials, Imperial College London, Exhibition Road, London SW7 2AZ, UK
[3] Department of Quantum Chemistry, St. Petersburg University, Universitetsky Prosp. 26, 198504 St. Petergof, Russia
[4] Institute for Solid State Physics, Kengaraga 8, LV-1063 Riga, Latvia

$La_{1-x}Sr_xCo_{0.25}Fe_{0.75}O_{3-\delta}$ is known as one of the best cathode materials for permeation membranes and solid oxide fuel cells. Optimization of its chemical composition is a challenging problem. One of the key properties is concentration of oxygen vacancies, controlled by their formation energies. *Ab initio* calculations were employed in order to study the formation of oxygen vacancies in $La_{1-x}Sr_xCo_{0.25}Fe_{0.75}O_{3-\delta}$ perovskites by varying the Sr content from x = 12.5% to 50%. The formation energies were obtained for different stoichiometries as functions of temperature and oxygen partial pressure. We have shown that the phonon contribution to the free formation energy becomes increasingly important in $La_{1-x}Sr_xCo_{0.25}Fe_{0.75}O_{3-\delta}$ not only with rising temperature but also with rising Sr content. We have shown that the formation energies are decreased significantly with increasing Sr content due to two effects: charge compensation of $Sr^{2+}$ ions and phonon contribution. We have suggested a simple explanation to increasing role of phonons for the oxygen vacancies formation energies on the basis of phonon mode changes in comparison to defect-free materials. A careful analysis of the experimental results from the literature is also presented.

**DOI**                                                                **PACS**

## 1. INTRODUCTION.

Cobaltite based electrode materials, like $(La,Sr)(Co,Fe)O_3$ are extensively studied for applications in oxygen permeation membranes [1] and fuel cells [2]. The quality of cathode material for fuel cells is to a large extent determined by its stability in oxidizing atmosphere and its level of catalytic activity for the oxygen reduction reaction [2]. The role of oxygen vacancies for these processes is crucial and its understanding requires analysis of their behavior on the basis of *ab initio* calculations.

Recently, complex perovskite oxides attracted great attention from chemists, who now have efficient computation methods at their disposal to calculate the electronic structure of such systems [3-9]. Many of these calculations deal with the behavior of point defects. However, the role of phonons is often overlooked or not carefully treated, which may lead to significant errors in point defect formation energies. Nowadays, the phonon properties of materials can be effectively calculated *ab initio* [10-12]. In the present study we have addressed this question together with symmetry aspects important for the choice of super-structures. The purpose of our paper is to show the effect of vibrational free energy, including the free energy of formation of oxygen vacancies for $(La,Sr)(Co,Fe)O_3$. We will explain how the lattice vibrations contribute to decreasing formation energies of oxygen vacancies with the Sr content and demonstrate the effect of phonon mode changes in the presence of oxygen vacancies.

Section I is devoted to symmetry considerations of the possible super-structures for $La_{1-x}Sr_xCo_{0.25}Fe_{0.75}O_{3-\delta}$ that we have treated. Section II describes main results for bulk properties and defects in $La_{1-x}Sr_xCo_{0.25}Fe_{0.75}O_{3-\delta}$. In Section III we present main conclusions.

## 2. STRUCTURE AND SYMMETRY OF MODELS.

The modeling of defective crystals for different defect concentrations is usually performed with periodic boundary conditions based on the supercell approach [13]. Let us suppose that Co atoms substitute for Fe in the cubic perovskite structure of $LaFeO_3$, and the atomic fraction of Co is 0.25. This requires a 2x2x2 extension of the primitive unit cell giving 40 atoms in the super-cell, i.e. 8 La, 8 Fe and 24 O atoms. There are three different ways to arrange two Co atoms (fig. 1) to occupy symmetry-equivalent Fe sites in this cubic structure. These three different arrangements of Co atoms (super-structures) will lead to different changes to the $La(Co_{0.25}Fe_{0.75})O_3$ symmetry: two super-structures have tetragonal symmetry (figs. 1(b) and 1(c)) whereas the third one has a body-centred cubic symmetry (fig. 1(a)). The super-structures differ by the distribution of atoms over the Wyckoff positions. In the present work we used the FINDSYM program of Hatch and Stokes [14] in order to identify the corresponding space group (table 1) and Wyckoff positions of all atoms in the super-structure. The sub-lattices of our primary interest are the La sub-lattice where the Sr atoms will substitute for La (in two concentrations: 12.5% and 50%) and the O sub-lattice where the vacancy will be considered. We will use the following abbreviation to refer to two compositions used in the present study: L87.5S12.5 for $La_{0.875}Sr_{0.125}Co_{0.25}Fe_{0.75}O_3$ and L50S50 for $La_{0.5}Sr_{0.5}Co_{0.25}Fe_{0.75}O_3$. It was already observed experimentally that the symmetry of $La_{1-x}Sr_xCo_{0.25}Fe_{0.75}O_{3-\delta}$ crystals changes with the Sr content [15-17]. The structural changes involve the symmetry reduction from rhombohedral to orthorhombic. At the same time, the cubic symmetry can also be observed for high concentrations of Sr accompanied by high deviations from stoichiometry on the oxygen sub-lattice. The symmetry changes are determined by the local site symmetry in the super-structure. All three super-structures in fig. 1 have one Wyckoff position for La atoms. However, when one Sr atom (the case of L87.5S12.5) substitutes for La (table 1) in the super-structure (a), the symmetry changes from cubic (Im-3m) to rhombohedral (R-3m); while in the super-structures (b) and (c) the symmetry changes from tetragonal (P4/mmm) to orthorhombic (Amm2 and Cmm2).

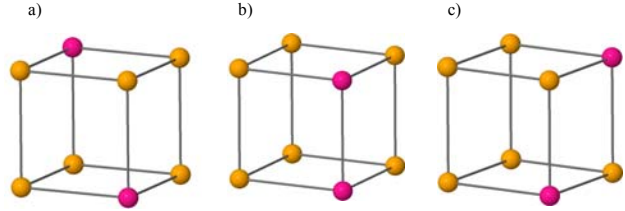

Fig. 1. The Fe atom sublattice in cubic 40 atoms supercell of $LaFeO_3$ with three different possibilities to arrange two Co atoms (pink atoms).

Table 1. Space groups of different super-structures as depicted in figs. 1 and 2 with Co and Sr atoms substitution in cubic perovskite structure. L87.5S12.5 and L50S50 correspond to $La_{0.875}Sr_{0.125}Co_{0.25}Fe_{0.75}O_3$ and $La_{0.5}Sr_{0.5}Co_{0.25}Fe_{0.75}O_3$, respectively.

| Super-Structure (fig. 1) | $LaCo_{0.25}Fe_{0.75}O_3$ | L87.5S12.5 | L50S50 (fig. 2) | | |
|---|---|---|---|---|---|
| | | | d | e | f |
| a | Im-3m | R-3m | Fmmm | P4/nmm | Pn-3m |
| b | P4/mmm | **Amm2** | Pmma | 1) P4/mmm 2) Pmm2 | $P4_2/mcm$ |
| c | P4/mmm | Cmm2 | Pmna | **1) Pmma** 2) P4mm | P-4m2 |

We start with the super-structures on the Co/Fe sub-lattice as depicted in fig. 1 and consider the substitution of La atoms by four Sr atoms (the case of L50S50). There are three possible arrangements for 4 Sr atoms in the cubic La sub-lattice. With the aim of studying the energetically preferable configuration for L50S50 we have to consider all three La-Sr arrangements for each super-structure in fig. 1. Fig. 2 illustrates the three arrangements for Sr atoms on the La/Sr sub-lattice. All space groups from combinations of super-structures in figs. 1-2 are collected in table 1. There are also possibilities to have two space groups for one combination (super-structure e in table 1). Two space groups are due to different orientations of the plane of Sr atoms in super-structure e with respect to the two Co atoms in super-structures b and c.

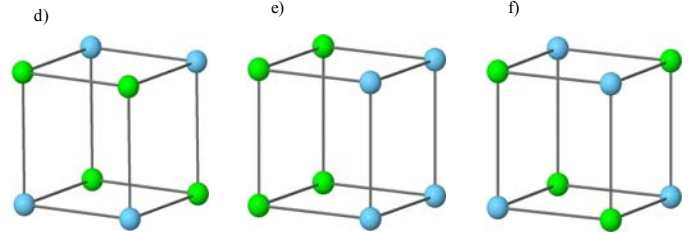

Fig. 2. The La atom sub-lattice in cubic 40 atoms supercell of $LaFeO_3$ with three different possibilities to arrange four Sr atoms (cyan atoms).

Having determined the optimum structure for fixed stoichiometries on the Co/Fe and La/Sr sub-lattices, the oxygen vacancy can be tackled. We denote the concentration per site of oxygen vacancies by $c_V$. In our calculations on oxygen vacancy in L50S50 and L87.5S12.5 only the super-cell of 40 atoms giving $c_V = 12.5\%$ was chosen. The oxygen vacancy represented an $F$-center (neutral super-cell). In the following we use Kröger-Vink notations [18] for the analysis of our results and denote the oxygen vacancy $V_o^{\bullet\bullet}$. In order to discuss all possible space groups for crystals with $V_o^{\bullet\bullet}$'s, let us mention that crystals with orthorhombic (Amm2) and tetragonal (Pmma) symmetry represent the ground state for L87.5S12.5 and L50S50, respectively (this will also be discussed quantitatively below). The 24 oxygen atoms are distributed over 10 and 8 Wyckoff positions in L87.5S12.5 and L50S50, respectively. The symmetry of such crystals with $V_o^{\bullet\bullet}$ changes in accordance with the location of the vacant site. The space groups of crystals with one $V_o^{\bullet\bullet}$ per super-cell are collected in table 2. There are three space groups for L87.5S12.5 and two space groups for L50S50 with $V_o^{\bullet\bullet}$. In case of Sr 50%, the oxygen vacancy prefers long distances from the plane of Sr atoms. However, the two super-structures (table 2) differ by the distance to Co atoms, too. It is at the shortest distance for the super-structure *b+e*. In case of Sr 12.5%, the oxygen vacancy prefers long distance from Co atoms but short distance to Sr atom.

Table 2. Space groups of L87.5S12.5 and L50S50 with one $V_o^{\bullet\bullet}$ per supercell at different Wyckoff positions. Only one super-structure *b* for L87.5S12.5, and two super-structures 1) *c* and *e*, 2) *b* and *e* for L50S50 are chosen for Co/Fe and Sr/La sub-lattices, respectively (see the text and table 1). The most energetically favourable structures are indicated with bold.

| Compound | Space group |
|---|---|
| L87.5S12.5: *b* | P1 |
|  | **Amm2** |
|  | Pm |
| L50S50: *c* + *e* | **Pmm2** |
|  | Pm |
| L50S50: *b* + *e* | **Pmm2** |
|  | Pm |

## 3. COMPUTATIONAL DETAILS.

In the present study the contribution of lattice vibrations to formation energies of oxygen vacancies in complex perovskite oxides is determined quantitatively. The super-cells for both L50S50 and L87.5S12.5 were chosen to contain 40 atoms. The phonon frequencies were calculated within the harmonic approximation as implemented in the VASP 5.2 code [19-20]. We used the projector augmented wave (PAW) method [21] and scalar-relativistic pseudopotentials (PPs) substituting for 46 electrons on La , 18 electrons on Co, 12 electrons on Fe, 28 electrons on Sr and 2 electrons on O atoms. The exchange-correlation functional was that of Perdew, Burke and Ernzerhof [22]. The plane wave cut-off energy was fixed at 520 eV for the geometry optimization and increased to 600 eV for the phonon frequency calculations. The lattice parameters and internal coordinates were optimized until the energy difference reached $10^{-5}$ eV. The integrations in reciprocal space over the Brillouin zone were performed using a 4x4x4 Monkhorst-Pack scheme [23]. The electron occupancies were determined with the Gaussian method [24] using a smearing parameter of 0.2 eV. A ferromagnetic structure of $La_{1-x}Sr_xCo_{0.25}Fe_{0.75}O_{3-\delta}$ was assumed throughout all the spin-polarized calculations.

## 4. RESULTS AND DISCUSSION.

### 4.1. Vibrational contribution to formation energies

In the present study we used the orthorhombic symmetry for both L50S50 (space group Pmma) and L87.5S12.5 (space group Amm2) crystals. The crystals with this symmetry have the lowest calculated total energies in comparison to crystals with other symmetries from table 1; the difference in total energy for L87.5S12.5 is 0.88(0.56) eV per (40 atoms) super-cell between the super-structure with Amm2 and R-3m(Cmm2) space groups. Orthorhombic symmetry is in accordance with the experiments of Tai et. al [15]. However, the situation with the Sr concentration of 50% is more complex. Tai et. al [15] and Swierczek et. al [17] demonstrated a monoclinic whereas Lein et. al [24] concluded a rhombohedral symmetry for $La_{0.5}Sr_{0.5}Co_{0.5}Fe_{0.5}O_{3-\delta}$. Hashimoto et. al [16] concluded from their XRD studies that the symmetry of $La_{0.6}Sr_{0.4}Co_{0.2}Fe_{0.8}O_{3-\delta}$ must be rhombohedral in air. We should not exclude the role of vacancies for these considerations. Monoclinic symmetry (space group Pm) is also observed in our study for one vacancy in both L50S50 and L87.5S12.5 (table 2). The difference in total energy between the super-structure with the Pmma space group and all the others from table 1 varies from 3.12 meV to 0.53 eV. The smallest difference in total energy is observed between the two symmetries for the combination of *c* and *e* super-structures (table 1). We will use also the combination of super-structures *b* and *e* for comparison.

The size of super-cell with 40 atoms ($\delta = 0.125$) is sufficient for L50S50 based on the experimental results of Tai et. al [15] and Lein et. al [25]. In these experiments the measured stoichiometry deviation was $\delta = 0.07$ for $La_{0.6}Sr_{0.4}Co_{0.2}Fe_{0.8}O_{3-\delta}$ and 0.14 for $La_{0.5}Sr_{0.5}Co_{0.5}Fe_{0.5}O_{3-\delta}$ at 1073 K and $10^{-1}$ bar. The oxygen vacancies in L50S50 may play an important role being at least partly responsible for the decrease in electrical conductivity at high temperatures as shown, for example, by Tai et. al [15] and Bouwmeester et. al [26]. In contrast, the compound L87.5S12.5 is highly stoichiometric within a broad temperature range in air [15]. Deviations from stoichiometry of the order of 2% at the maximum are only expected at high temperatures. Similar stoichiometry deviations were seen in the experiments of Mizusaki et. al for $La_{0.9}Sr_{0.1}CoO_3$ [27] under the same conditions. These experimental results, thus, indicate higher formation energies of $V_o^{\bullet\bullet}$ for L87.5S12.5. Below we will discuss the importance of these aspects for the formation energy of $V_o^{\bullet\bullet}$.

The $Sr^{2+}$ ion substituting for $La^{3+}$ must be compensated by electronic defects (holes) or $V_o^{\bullet\bullet}$'s. The oxygen vacancies prevail over electronic defects at high temperatures [15, 26]. We calculated the free formation energy of $V_o^{\bullet\bullet}$ as a function of temperature following the expression:

$$\Delta G_F(T, p_{O_2}) = \left[ E_{tot}^{V_O^{\bullet\bullet}} + \frac{1}{n}\left(E_{vib}^{V_O^{\bullet\bullet}} - TS_{vib}^{V_O^{\bullet\bullet}}\right) + PV^{V_O^{\bullet\bullet}} \right] - \left[ E_{tot}^{p} + \frac{1}{n}\left(E_{vib}^{p} - TS_{vib}^{p}\right) + PV^{p} \right] + \mu_o(T, p_{O_2}),$$ (1)

where superscripts $p$, $V_O^{\bullet\bullet}$ stand for pure and defective crystals, $E_{tot}$ total static internal energies, $E_{vib}$ the vibrational contribution, including zero point motion to internal energy, $S_{vib}$ the entropy of vibrations, $V$ the super-cell volume, $T$ the temperature, and $P$ the standard pressure. The oxygen chemical potential $\mu_o(T, p_{O_2})$ was estimated following the approach discussed elsewhere [28-29], *viz.*

$$\mu_o(T, p_{O_2}) = E_{tot}^{AO} + \frac{1}{m}\left(E_{vib}^{AO}(T^0) - TS_{vib}^{AO}(T^0)\right) - E_{tot}^{A} - \frac{1}{l}\left(E_{vib}^{A}(T^0) - TS_{vib}^{A}(T^0)\right) - \Delta G^{AO}(T^0) + \Delta\mu_o(T) + \frac{1}{2}k_B T \ln\left(\frac{p_{O_2}}{p^0}\right),$$ (2)

where superscripts $AO$, $A$ stand for a binary oxide and corresponding metal, $T^0$ an arbitrary reference temperature which is equal in our case to room temperature, $\Delta G^{AO}(T^0)$ (= 601.6 kJ/mol) the binary oxide formation energy at standard temperature taken from a thermodynamic database [30], $\Delta\mu_o(T)$ is the difference in chemical potentials at the temperature of interest and reference temperature, which is taken from the thermodynamic database [30], $k_B$ Boltzmann's constant, $p_{O_2}$ and $p^0$ are the oxygen partial pressure and standard pressure, respectively.

The standard expressions for $E_{vib}$ and $S_{vib}$ within the harmonic approximation are discussed in many reviews (see, [11] and references therein). The calculation of enthalpy of $V_O^{\bullet\bullet}$ formation $\Delta H_F$ corresponds to neglecting all entropy $TS$ and $\frac{1}{2}k_B T \ln\left(\frac{p_{O_2}}{p^0}\right)$ terms in Eqs. 1-2 provided the zero-point vibrations are included in $E_{vib}$. Note that Eq. 2 is based on the assumption that $\mu_{AO} = \mu_A + \mu_O$. The pre-factors 1/n, 1/l, 1/m in Eqs. 1-2 represent the ratios of the number of atoms in the primitive unit cell to that in the super-cell for defective (perfect) $La_{1-x}Sr_xCo_{0.25}Fe_{0.75}O_3$, binary oxide and metal, respectively. Thus, $n = 1$, $m = l = 27$ (3x3x3 extension of hexagonal and cubic primitive unit cells of Mg and MgO, respectively) in our calculations. We have chosen MgO as a reliable reference material because it is relatively easy to calculate *ab initio*. Note that the total energy and the vibrational contributions from corresponding binary oxide and metal are used in order to avoid calculations of $O_2$ molecule properties in the PW basis with the PBE functional. In this case the $O_2$ molecular binding energy deviates significantly from the experimental value (by > 1.5 eV) [31] which could be a source of error for the formation energy of $V_O^{\bullet\bullet}$. Recently, we showed [32] that properties of $O_2$ such as equilibrium distance, binding energy and vibrational frequency are successfully reproduced with the hybrid PBE0 functional [33] in the LCAO basis. It is worth mentioning that the impact of phonons for the formation energy discussed below does not change with the choice of other binary oxides in Eq. 2. Besides, Fig. 3 demonstrates a comparison of phonon contribution to the chemical potential $\mu_O$ calculated employing hybrid PBE0 (adopted from Ref. 32) and Eq. 2. The two approaches are well compared with largest difference between them being 0.03 eV. The phonon contribution to $\Delta G_F$ from AO and A in Eq. 2 at room temperature is

$$E_{vib}^{AO}(T^0) - TS_{vib}^{AO}(T^0) - E_{vib}^{A}(T^0) + TS_{vib}^{A}(T^0) \approx 0.14 \text{ eV}$$

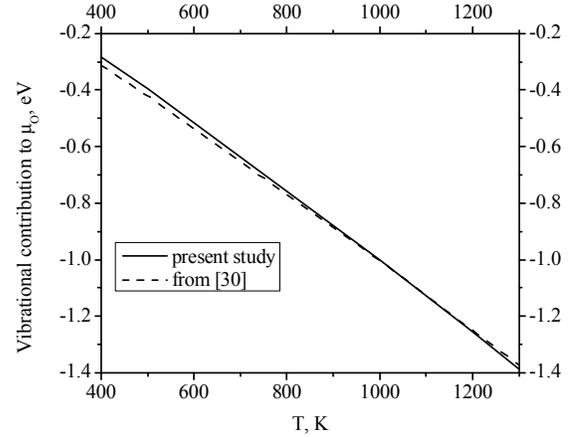

Fig. 3. A comparison of vibrational contribution to $\mu_O$ calculated by two methods: hybrid functional from Ref. 32 and Eq. 2 in the present study.

In fig. 4a the calculated free formation energy $\Delta G_F$ of $V_O^{\bullet\bullet}$ at standard oxygen pressure is given for both L50S50 (red curves) and L87.5S12.5 (black curves). The role of phonons is also demonstrated by calculating $\Delta G_F$ without the phonon contribution. The phonon contribution increases with T for both compounds. As expected $\Delta G_F$ is smaller for L50S50 in comparison to L87.5S12.5 in accordance with experimental findings. Moreover, the difference between $\Delta G_F$'s of the two compounds increases with $T$ from 1.4 eV at 400 K to 1.9 eV at 1300 K. If, however, the phonon contribution is neglected, then the temperature dependence is given by the chemical potential of oxygen only, and the difference between $\Delta G_F$'s of the two compounds remains constant at all T (dashed curves). As also seen in fig. 4a the role of phonons is different in the two compounds, which is reflected in increased (decreased) $\Delta G_F$ in L87.5S12.5 (L50S50). Note that $\Delta G_F$ in fig. 4a was calculated for the case of L50S50:$c+e$. The effect of phonons on $\Delta G_F$ may be different for other vacancies and/or super-structures. For example, the case of L50S50:*b+e* is characterized by a $\Delta G_F$-increase due to phonons (fig. 5) similar to L87.5S12.5. Also, $\Delta G_F$ for L50S50:*b+e* is close to that for L50S50:*c+e* if the phonon contribution is neglected. We, thus, suggest that the proper choice of super-structures and inclusion of phonons are important to calculate the formation energies of $V_O^{\bullet\bullet}$'s in complex oxides. The free formation energy with phonons for the two combinations in fig. 4 differ by almost 1 eV at T = 1300 K. To better understand the effect of phonons for $\Delta G_F$ in the two compounds and two combinations in figs. 4-5, one could consider that the phonon contribution to $\Delta G_F$ is mainly given by

$$\left(E_{vib}^{V_O^{\bullet\bullet}} - TS_{vib}^{V_O^{\bullet\bullet}}\right) - \left(E_{vib}^{p} - TS_{vib}^{p}\right) + vibrational\ contribution\ to\ \mu_o.$$

Since the vibrational part to $\mu_O$ is always the same, let us consider for simplicity the difference $\Delta = \left(E_{vib}^{V_O^{\bullet\bullet}} - TS_{vib}^{V_O^{\bullet\bullet}}\right) - \left(E_{vib}^{p} - TS_{vib}^{p}\right)$ only. In fig. 6 the difference is plotted as a function of T. It increases and remains positive for L87.5S12.5 and L50S50:*b+e* at T>500 K and T>800 K, respectively. It is, however, negative for L50S50:*c+e* at all T. We consider this effect as a very good indication of important role of $V_O^{\bullet\bullet}$ in L50S50. Such a behavior of $\Delta$ as shown in fig. 6 is possible if the frequencies in the presence of $V_O^{\bullet\bullet}$ become softer (the case of L50S50:*c+e*) or harder (the case of L87.5S12.5 and L50S50:*b+e*) in comparison to the pure bulk material especially in the region of low frequencies. In fig. 7 the total phonon density of states is presented for all the three cases considered here. The L50S50:*c+e* compound has an increased phonon density in the range from 0 to 50 cm$^{-1}$ in comparison to

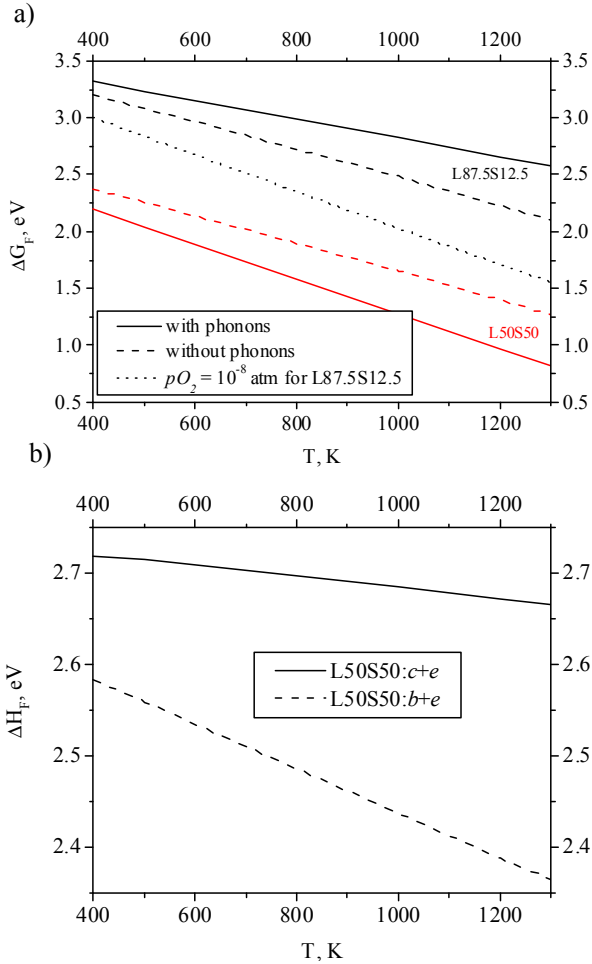

Fig. 4. a) The free formation energy $\Delta G_F$ per unit cell of $V_O^{\bullet\bullet}$ as a function of temperature at standard oxygen pressure for L87.5S12.5 (solid and dashed black curves) and L50S50 (solid and dashed red curves, the combination of super-structures *c* and *e*). The dotted curve represents $\Delta G_F$ including the phonon contribution for L87.5S12.5 at oxygen partial pressure $p_{O_2} = 10^{-8}$ atm. b) The enthalpy of $V_O^{\bullet\bullet}$ formation for L50S50 as a function of temperature with phonon contribution included.

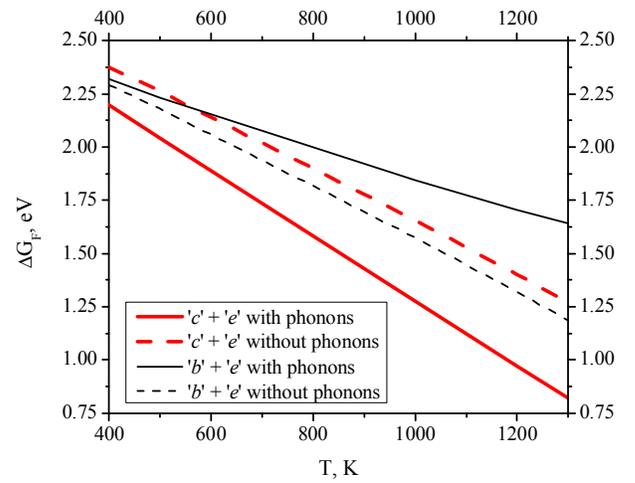

Fig. 5. The free formation energy per unit cell of $V_O^{\bullet\bullet}$ as a function of temperature at standard pressure for two combinations of super-structures for L50S50 (bold and red curves for *c* and *e*, black curves for *b* and *e*).

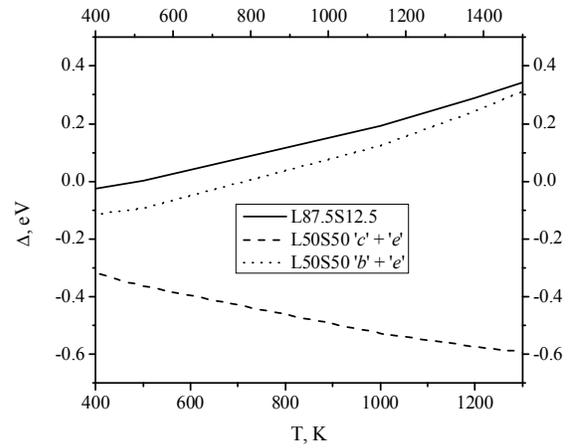

Fig. 6. The energy difference $\Delta = \left(E_{vib}^{V_O^{\bullet\bullet}} - TS_{vib}^{V_O^{\bullet\bullet}}\right) - \left(E_{vib}^{p} - TS_{vib}^{p}\right)$ as a function of T for L87.5S12.5 and L50S50.

L87.5S12.5 and L50S50:$b+e$. Specifically in this range the softer phonons have significant contribution to the free energy, and, if they are lower than in the pure bulk material, then $\Delta$ will be negative. The effect of soft frequencies is particularly important for high temperatures. According to our results, the contribution of lowest frequency (25 cm$^{-1}$) to the free energy is 35 times the contribution of highest frequency (612 cm$^{-1}$) at T = 1000 K for L50S50:$c+e$ with $V_O^{\bullet\bullet}$. The main contribution to soft vibrational modes is due to O, La and Sr atoms. The Fe and Co atoms contribute at higher frequencies whereas the O atoms are involved at all frequencies.

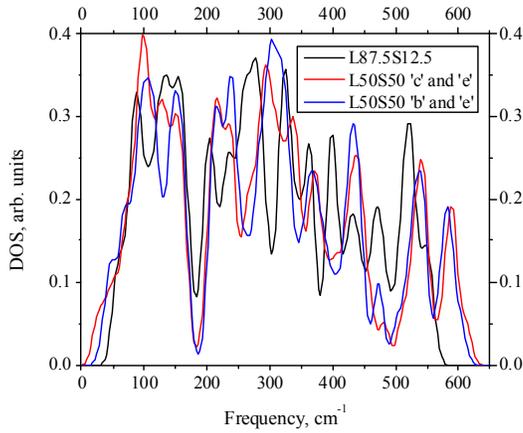

Fig. 7. The total phonon density of states (DOS) for L87.5S12.5 and L50S50 with $V_O^{\bullet\bullet}$.

### 4.2. A comparison to the experimental literature.

Our estimations of $\Delta G_F$ also included oxygen partial pressure for L87.5S12.5 (fig. 4a, dotted curve). According to the measurements [26] very low $p_{O_2}$ (of the order of 10$^{-8}$ atm) would be needed for the calculated vacancy concentration of $c_V$=12.5%. The calculated $\Delta G_F$ then decreases from 2.82 eV (standard pressure) to 2.03 eV ($p_{O_2}$ = 10$^{-8}$ atm) at T = 1000 K. On the other hand, a better comparison to the experiment value can be provided by comparing the enthalpies of formation. Our comparison to the experimental literature included thermogravimetry measurements of Mizusaki et. al on (La,Sr)CoO$_3$ [27] and (La,Sr)FeO$_3$ [34]. In these measurements the partial molar enthalpy of oxygen $\Delta h_o$ as a function of $\delta$ for different Sr concentrations was determined. This measured quantity is then equivalent to our calculated $\Delta H_F$. Thus, $\Delta h_o$ was 3.2 eV and 0.9 eV for La$_{0.9}$Sr$_{0.1}$CoO$_{3-\delta}$ and La$_{0.5}$Sr$_{0.5}$CoO$_{3-\delta}$, respectively. For La$_{0.9}$Sr$_{0.1}$FeO$_{3-\delta}$ $\Delta h_o$ varies almost step-wise between the oxidation states (4+/3+ amd 3+/2+ region). So, the values of 5.0 and 1.2 eV [34] correspond to respective formation enthalpies. The calculated values of $\Delta H_F$ slightly depend on the temperature and suggest 3.7 eV for L87.5S12.5 at high temperatures. This calculated value compares fairly well to the average values for La$_{0.9}$Sr$_{0.1}$FeO$_{3-\delta}$. Contrarily, the calculated $\Delta H_F$ (=2.6 eV, i.e. average value in fig. 4b) for L50S50 is by 1.7 eV larger than the measured one for La$_{0.5}$Sr$_{0.5}$CoO$_{3-\delta}$ [27]. The respective $\Delta h_o$ for La$_{0.4}$Sr$_{0.6}$FeO$_{3-\delta}$ and La$_{0.6}$Sr$_{0.4}$FeO$_{3-\delta}$ amounts to ≈ 1 eV. A similar oxidation enthalpy was measured by Bucher et al. [35] for La$_{0.4}$Sr$_{0.6}$Fe$_{0.4}$Co$_{0.6}$O$_{3-\delta}$. As seen in fig. 4b the formation enthalpy for the two cases L50S50:$b+e$ and L50S50:$c+e$ differ by 0.5 eV at high temperatures. Note that we expect the formation enthalpy and free formation energy to decrease by 0.3 eV when the super-cell size changes from 40 atoms to 320 atoms as it was shown by Kotomin et. al [5] in their calculations on Ba$_{0.5}$Sr$_{0.5}$Co$_{0.75}$Sr$_{0.25}$O$_{3-\delta}$.

Note that we have adopted the specific permeability measurements to further compare the calculated formation energies to experiments which will be discussed in detail in our future paper [36].

### 5. CONCLUSIONS

In the present study we have shown that the lattice vibrations (phonons) play a significant role in determining the equilibrium concentration of oxygen vacancies in La$_{1-x}$Sr$_x$Co$_{0.25}$Fe$_{0.75}$O$_3$, which is the more important the higher the temperature. This is very much in contrast to SrTiO$_3$ where the main temperature contribution to the oxygen vacancy formation energy is due to chemical potential of oxygen [31]. Such results suggest than not only charge compensation but also the lattice vibrations are responsible for higher concentrations of oxygen vacancies in La$_{1-x}$Sr$_x$Co$_{0.25}$Fe$_{0.75}$O$_3$ with higher concentrations of Sr. We have demonstrated quantitatively that the phonons should not be neglected when calculating oxygen vacancy formation energies as a function of temperature in complex oxides especially with high dopant concentrations. Moreover, the role of phonons in the defect formation energy depends greatly on the oxygen vacancy position in the perovskite structure. Depending on the particular position, the low frequency phonon modes can soften or harden in comparison to the defect-free material. We, thus, emphasize that to calculate properly the formation energies of oxygen vacancies and their concentrations, one needs to carefully consider all possible defect arrangements and take into account the phonon contribution.

### ACKNOWLEDGEMENTS

The research leading to these results received partial support from the FP7 NASA-OTM project (grant NMP3-SL-2009-228701). D.G. thanks HPC-Europa2 initiative of European Commission and Thomas Young Centre for financial support of his visit to Imperial College London. The authors thank the IT team of EPCC in Edinburgh for technical support and E. Blokhin for help with plotting the graphs. Authors are grateful to Rotraut Merkle and Eugene Heifets for helpful discussions.

* d.gryaznov@fkf.mpg.de